\documentclass{article}

\usepackage{arxiv}
\usepackage[utf8]{inputenc} 
\usepackage[T1]{fontenc}    
\usepackage{hyperref}       
\usepackage{url}            
\usepackage{booktabs}       
\usepackage{amsfonts}       
\usepackage{microtype}      
\usepackage{lipsum}		
\usepackage{graphicx}
\usepackage{doi}
\usepackage{siunitx}
\usepackage{amsmath}
\usepackage{paralist}
\usepackage{setspace}

\usepackage[font=large]{caption}

\title{Low-symmetry non-local transport in microstructured squares of delafossite metals}

\date{}

\hypersetup{
pdftitle={A template for the arxiv style},
pdfsubject={q-bio.NC, q-bio.QM},
pdfauthor={David S.~Hippocampus, Elias D.~Striatum},
pdfkeywords={First keyword, Second keyword, More},
}

\begin{document}
\large 
\maketitle
\vspace{-1cm}

\setstretch{1.1}

\begin{center}

Philippa H. McGuinness$^{1,2\dag}$, Elina Zhakina$^{1,2\dag}$, Markus König$^{1}$, Maja D. Bachmann$^{1,2}$, Carsten Putzke$^{1,3}$, Philip J.W. Moll$^{1,3}$, Seunghyun Khim$^{1}$ and Andrew P. Mackenzie$^{1,2*}$\\ 
\end{center}

\large 
\begin{enumerate}[$^1$]
\item Max Planck Institute for Chemical Physics of Solids, Nöthnitzer Str. 40, 01187 Dresden, Germany
\item Scottish Universities Physics Alliance, School of Physics \& Astronomy, University of St. Andrews, North Haugh, St. Andrews KY16 9SS, UK
\item Institute of Materials, École Polytechnique Fédérale de Lausanne (EPFL), 1015 Lausanne, Switzerland\\

$^\dag$These authors contributed equally. $^*$andy.mackenzie@cpfs.mpg.de 
\end{enumerate}

\begin{abstract}
Intense work studying the ballistic regime of electron transport in two dimensional systems based on semiconductors and graphene had been thought to have established most of the key experimental facts of the field.  In recent years, however, new forms of ballistic transport have become accessible in the quasi-two-dimensional delafossite metals, whose Fermi wavelength is a factor of 100 shorter than those typically studied in the previous work, and whose Fermi surfaces are nearly hexagonal in shape, and therefore strongly faceted.  This has some profound consequences for results obtained from the classic ballistic transport experiment of studying bend and Hall resistances in mesoscopic squares fabricated from delafossite single crystals.  We observe pronounced anisotropies in bend resistances and even a Hall voltage that is strongly asymmetric in magnetic field.  Although some of our observations are non-intuitive at first sight, we show that they can be understood within a non-local Landauer-Büttiker analysis tailored to the symmetries of the square/hexagonal geometries of our combined device/Fermi surface system.  Signatures of non-local transport can be resolved for squares of linear dimension of nearly 100~\si{\micro}m, approximately a factor of 15 larger than the bulk mean free path of the crystal from which the device was fabricated.  
\end{abstract}


\section{Introduction}

The ballistic regime, in which the electron mean free path is larger than a characteristic geometric length scale, is an unconventional regime of non-local electrical transport which only occurs within ultrapure materials. A common method to examine materials in this regime is to study electrical transport in four-terminal junctions which are smaller than the electron mean free path. Within cross or square shaped devices, unconventional effects can occur, such as a negative non-local ‘bend’ resistance \cite{hirayama_hall_1991,timp_propagation_1988,takagaki_nonlocal_1989} and an enhanced, suppressed or even negative Hall resistance \cite{ford_vanishing_1988,ford_influence_1989,gilbertson_ballistic_2011,roukes_quenching_1987}. Estimates of the characteristic length scales of the ballistic regime can also be made by examining the suppression of these effects with increasing device size~\cite{matioli_room-temperature_2015,tarucha_bend-resistance_1992,hirayama_ballistic_1991,sakamoto_impurity_1991}.

The majority of four-terminal ballistic regime studies, however, have concentrated on  semiconductor heterostructures~\cite{daumer_quasiballistic_2003-1,goel_ballistic_2005,gilbertson_ballistic_2011,matioli_room-temperature_2015} or monolayer graphene~\cite{banszerus_ballistic_2016,bock_influence_2012,mayorov_micrometer-scale_2011,weingart_low-temperature_2009} rather than metals, primarily because the electronic mean free path in quasi-two-dimensional metals is typically less than 100 nm. In recent years, however, it has become clear that the delafossite metals PdCoO$_2$ and PtCoO$_2$~\cite{shannon_chemistry_1971-2,mackenzie_properties_2017,takatsu_roles_2007} have the potential to be ideal hosts for the study of non-local transport effects. Almost uniquely among oxide metals, they have an extremely high purity as grown, with defect levels as low as 1 in 10$^5$ in the conducting Pd/Pt layers~\cite{sunko_controlled_2020}. They also bring the benefit of simplicity, because only one highly dispersive band crosses the Fermi level, giving a single, quasi-two-dimensional Fermi surface as established experimentally by angle-resolved photoemission and measurements of the de Haas-van Alphen effect~\cite{arnold_fermi_2020,hicks_quantum_2012,kushwaha_nearly_2015,noh_orbital_2009}. Due to the high purity, momentum relaxing mean free paths as long as 20~\si{\micro}m can be achieved at low temperature ~\cite{hicks_quantum_2012,takatsu_extremely_2013,kushwaha_nearly_2015,mackenzie_properties_2017}, enabling the observation of multiple unconventional transport and quantum interference effects in mesoscopic samples~\cite{bachmann_super-geometric_2019,moll_evidence_2016,putzke_h_2020}. 

For various non-local in-plane transport properties, the shape of the Fermi surface of PdCoO$_2$ has been demonstrated to strongly influence observations.  In particular, its nearly hexagonal cross-section leads to a high level of directionality of transport properties once the non-local regime is entered at low temperatures~\cite{bachmann_directional_2021,cook_electron_2019}. This pronounced Fermi surface faceting is one way in which non-local transport measurements on delafossite metals differ from those on semiconductor two-dimensional electron gases or monolayer graphene, for both of which Fermi surfaces are close to circular. A second noteworthy feature is that, since the carrier density per layer in the delafossites is metallic (\num{1.5e15}\ cm$^{-2}$), the Fermi wavelength is two orders of magnitude shorter than that of typical doped semiconductors, placing the delafossites in a new regime of two-dimensional transport.  For these reasons it is interesting to make contact between the two fields by performing non-local transport measurements in some of the ‘classic’ device geometries of mesoscopic physics, to investigate similarities and differences between the behavior of the delafossite metals and the previously studied low carrier density systems.  In this paper we report on experiments designed to do this, studying the bend and Hall resistances of four-terminal microstructured squares.

\section{Results}
\label{sec:Results}
\subsection{Device fabrication}
We fabricated our square devices from crystals of the ultrapure metallic delafossites PtCoO$_2$  and PdCoO$_2$ grown in quartz tubes via methods discussed in~\cite{kushwaha_nearly_2015,kushwaha_single_2017}. The average low temperature mean free path within the crystals studied was around 5 \si{\micro} m for PtCoO$_2$ and 20 \si{\micro} m for PdCoO$_2$, highlighting their extreme purity. Due to the non-circular Fermi surface, which, importantly, belongs to a higher order D$_6$ symmetry group than the D$_4$ of the square device, the crystal orientation relative to the square becomes a potentially important parameter. The hexagonal symmetry ensures that there are only six predominant electron directions compared to a continuum for a circular Fermi surface, as illustrated in Figs. 1a-c. 

In the previously reported work on semiconductors, transmission of electrons along the square diagonal led to electrical transport features considered to be indicative of the ballistic regime. We therefore constructed squares with two different Fermi surface orientations: the ‘enhanced’ orientation, Fig. 1b, in which two of the six electron directions are parallel to the square diagonal, enhancing transmission along this diagonal compared to a circular Fermi surface, and the ‘diminished’ orientation, Fig. 1c, in which none of the directions are parallel to the diagonal, reducing diagonal transmission compared to a circular Fermi surface. In both orientations, the combination of the square and Fermi surface has D$_2$ symmetry, but only in the enhanced orientation are two of the mirror planes of the Fermi surface aligned along the square diagonals. In the diminished orientation, there is no symmetry along the square diagonals, with two of the mirror planes instead being horizontal and vertical and therefore aligned parallel to the square edges.

In total, we fabricated six PtCoO$_2$ squares, one with the enhanced orientation and five with the diminished orientation, and one PdCoO$_2$ diminished-orientation square using a focused ion beam (FIB) and standard microstructuring techniques similar to those described in~\cite{bachmann_manipulating_2019,moll_evidence_2016,moll_focused_2018}. An advantage of using FIB-based microstructuring is that, by alternating electrical transport measurements and further FIB steps to reduce the dimensions, many different square sizes can be studied using the same contacts and section of crystal. This reduces uncertainty introduced by a variation in purity within or between crystals. The initial side lengths of the fabricated squares ranged between 40 and 95~\si{\micro} m, with final values being as small as 7~\si{\micro} m. We chose to concentrate our studies primarily on PtCoO$_2$ rather than PdCoO$_2$ as the shorter mean free path enables a larger ratio of the square size to the mean free path to be reached within the limited available crystal size. Figs. 1d and e are scanning electron microscope images of a PtCoO$_2$ device before and after several cycles of FIB processing to reduce the dimensions. Vestiges of several intermediate square sizes can be seen from the trenches in the epoxy surrounding the smaller square.

\subsection{Bend voltage}

Inspired by the van der Pauw and Montgomery methods used to measure resistivity anisotropy in the Ohmic regime~\cite{montgomery_method_1971,van_der_pauw_method_1958} we began by performing voltage measurements with two different contact configurations, as displayed in Fig. 2a, which are related by a 90\si{\degree} rotation of all contacts. These are conventionally referred to as bend voltage measurements, as the current path bends around a corner in a cross device. Although the geometry is different, the nomenclature is also used for squares. They can be labelled as $V_{ij,kl}$ where the current flows amongst two adjacent contacts, i and j, and the voltage difference $V=V_k-V_l$ is measured between the other contacts. In the Bend 1 configuration ($V_{\mathrm{B}1} = V_{12,43}$) the current contacts, and hence also the voltage contacts, are linked by a horizontal translation whereas for the Bend 2 measurement ($V_{{B}2} = V_{23,14}$) they are linked vertically. 

For delafossite metals in the Ohmic regime, where the mean free path is much smaller than the device size, the underlying triangular lattice symmetry ensures an isotropic in-plane resistivity.  The bend voltages are therefore identical, and follow the van der Pauw equation~\cite{van_der_pauw_method_1958}.

\begin{equation}
V_{\mathrm{B}1,\mathrm{B}2}=\frac{\ln (2) \rho I}{t \pi}
\end{equation}
where $I$ is the applied current, $\rho$ is the material resistivity and $t$ is the thickness of the device, between 1 and 5 \si{\micro}m in our squares.  Importantly, this quantity is always positive and does not depend on the square side length. Across all our PtCoO$_2$ squares, the 300~K resistivity measured by this van der Pauw technique is $\rho$(300~K)~=~1.8~$\pm$~0.1 \si{\micro\Omega}cm, in good agreement with the accepted value of 1.82~\si{\micro\Omega}cm~\cite{nandi_unconventional_2018}.

Equation 1 does not remain valid within the ballistic regime due to its highly non-local nature and, in other materials, the bend voltage becomes negative in this regime, as the majority of injected electrons pass along the square diagonal to the negative voltage contact. However, no dependence of the bend voltage on the orientation of the contacts has been reported in any previous study of ballistic transport in four terminal devices: $V_{\mathrm{B}1}$ and $V_{\mathrm{B}2}$ were always identical.

In Fig. 2 we show the variation of $V_{\mathrm{B}1,\mathrm{B}2}t/I$ with temperature for two different PtCoO$_2$ squares, the enhanced-orientation E1 and the diminished-orientation D1. The voltage was measured using standard lock-in techniques at a frequency of 123~Hz. Typical currents were between 1 and 9~mA, although the voltage was linear over a current range of at least 0.09~mA to 9~mA (Fig. S1).  For each square, several cycles occurred in which the square size was reduced using the FIB and then the bend voltage measurements repeated, resulting in the range of side lengths shown.

At temperatures above 80~K, there is little dependence of the voltage on the size of the square, the orientation of the Fermi surface or the choice of bend measurement, as expected within the Ohmic regime and Equation 1. At low temperature, however, the voltage is strongly affected by all three of these quantities. 

In the enhanced orientation, Fig.~2b, at low temperature both bend voltages decrease rapidly as the square size decreases, eventually becoming negative at a square size of 15~\si{\micro}m and below. The two bend measurements are very similar, as expected from the symmetry of the Fermi surface about the diagonal shown in Fig.~2a, with only a small difference due to slight fabrication asymmetries. These observations are qualitatively consistent with what was found in semiconductor devices with circular Fermi surfaces~\cite{hirayama_ballistic_1991,goel_ballistic_2005}. 

However, other features of our measurements differ starkly from what was previously known. In particular, the behavior of the Bend 1 voltage profoundly changes between the Fermi surface orientations: in the diminished orientation square, Fig.~2d, this voltage rapidly increases as the square size decreases, opposite to the enhanced-orientation behavior. In contrast, the Bend 2 voltage decreases with square size, similar to the behavior in the enhanced orientation.  This anisotropy between the two bend measurements has never been observed before, and indeed cannot be observed in materials with a circular Fermi surface. The scale of the effect is also significant: in the 7~\si{\micro}m square, the difference between the bend resistances is as large as the total ohmic resistivity at 90~K. 

A strong dissimilarity is also seen in the magnetotransport of the two devices at 5~K, as shown in Fig. 3. Cyclotron orbits in these delafossites are hexagonal, but the cyclotron radius varies by less than 10\% from its average value across the whole orbit, so this value is a useful length scale for the problem. We define the average cyclotron radius $r_c = \hbar k_F/eB$, where $\hbar$ is Planck’s constant divided by 2$\pi$, $k_F$ the average Fermi wave vector, $e$ the electronic charge and $B$ the magnetic field. The dimensionless ratio $w/r_c$ is therefore proportional to magnetic field and often determines the scale of magnetoresistance effects in the ballistic regime, so we use it for the \emph{x}-axis of Fig. 3.

In the enhanced orientation, Fig.~3b, there is a dip at low field and a negative zero field voltage in small squares for both bend measurements. In a finite magnetic field, the electron trajectories are curved, which decreases the fraction of electrons passing along the square diagonal and increases the voltage. The width of the central dip as a function of $w/r_c$ is constant across multiple square sizes. At $w/r_c\sim 2$, there is a peak in the voltage, most visible at the smallest square sizes. This peak, and the harmonic at $w/r_c\sim 4$, are likely to be due to resonances similar to transverse electron focusing, where the cyclotron diameter and $w$ are commensurate and the majority of the electron trajectories are focused into an adjacent contact by the field~\cite{tsoi1974focusing,taychatanapat_electrically_2013} An enhancement of this effect due to the hexagonal Fermi surface has been previously demonstrated in delafossite metals~\cite{bachmann_super-geometric_2019}.

For the diminished orientation square, as shown in Fig.~3a, the Bend 2 voltage follows similar behavior to that of the enhanced orientation, with a dip at zero field and a peak around $2w/r_c$ and at multiples of this value. For the Bend 1 measurements, however, there is a peak at zero field, signifying the strong transport anisotropy present only within the ballistic regime. There are two small peaks located around $w/r_c = 0.12$ in the smallest square sizes. Since these peaks occur at the same $w/r_c$ independent of square size, they clearly stem from a geometric resonance between the square geometry and $r_c$. Although the physical origin of the resonances is likely similar to that of the focusing peaks, detailed simulations beyond the scope of this paper would be required to identify the precise electron trajectories involved. The anisotropy rapidly weakens as the square size is increased, but, surprisingly, there is still a visible peak for a 20~\si{\micro}m square, which has a side length around four times the typical mean free path. Data from even larger square sizes is also shown for comparison purposes but will be discussed in more detail later. In neither square is evidence observed of the fluctuations expected from the presence of quantum interference effects previously seen in semiconductors~\cite{takagaki_nonlocal_1989,baranger_classical_1991}. 

At the largest values of~$w/r_c$, the dependence of the voltage on the square size, the bend measurement and the Fermi surface orientation is weak. The small cyclotron radius at these field values ensures that the dominant electron trajectories are ‘guiding’ trajectories, which involve repeatedly scattering along the square boundary and eventually reaching an adjacent contact regardless of the square size or Fermi surface orientation.

The observation that significant bend voltage anisotropy still exists at a square size which is several multiples of the mean free path motivated us to examine the behavior at larger length scales in more detail.  In Fig.~3c we show the magnetotransport in a diminished orientation square, D2, for only the largest side lengths in Fig. 3a, between 40 and 95~\si{\micro}m. A distinct peak and dip remain present only up to 50~\si{\micro}m, but even at 95~\si{\micro}m there is a clear difference at zero field between the two bend voltages. The mean free path of this square was 6~\si{\micro}m, so the largest ratio of square width to mean free path is over 15:1, far outside the ratio of 1:1 that would signify the most strictly defined limit of the ballistic regime.

\subsection{Decay of anisotropy in bend voltage}
The limits of this ballistic behavior can be better characterized by quantifying how the anisotropy decays as a function of square size. In previous studies, the depth of the dip in the magnetoresistance at zero field has been used as a gauge of the strength of the ballistic effects~\cite{hirayama_ballistic_1991,tarucha_bend-resistance_1992,hirayama_high_1993}. Here, we use the zero-field difference at 5~K between the bend voltages in the diminished-orientation squares, $\Delta Vt/I = t[V_{\mathrm{B1}}(B = 0) - V_{\mathrm{B2}}(B = 0)]/I$. This parameter has the advantage of being zero within the Ohmic regime, as the bend resistances must be equal in this regime by symmetry, ensuring automatic subtraction of any Ohmic background. 

The variation of this quantity as a function of $w/l$, where $l$ is the mean free path of each square, is shown in Fig. 3d. With this choice of parameter, the effects of any variance in purity, and therefore $l$, between different devices,which causes different decay rates as a function of $w$ alone, is eliminated. To determine $l$ without influence from ballistic effects, the resistivity at high $w/r_c$, where these effects are suppressed, was calculated using Equation 1. The known bulk magnetoresistance~\cite{nandi_unconventional_2018} was used to compute the zero field resistivity, with $l$ then determined via standard Ohmic regime equations.
 
As demonstrated in Fig.~3d, there is good agreement between the decay and the exponential fit $\Delta Vt/I = Ae^{-b w/l}$  where $A$ and $b$ are fitted constants, providing that the decay constant $b$ has two different values, $b_R$ and $b_S$, in two regions of $w/l$. When $w/l$ < 5 , there is a rapid decay in $\Delta V$ as the square size is increased, with $b_R$ = 0.59~$\pm$~0.05. This is evidence of the expected strong dependence of these ballistic effects on the sample geometry. The decay rate is consistent between different squares. When $w/l$~>~5, there is a slower decay with $b_S = 0.21 \pm 0.02$, around a factor of three smaller than $b_R$. Intriguingly, we can track this much weaker ballistic decay to ratios of $w/l$ above 20. Even the rapid decay, however, is slower than those reported in semiconductor cross and square devices, which report single decays with equivalent  values of $b$ ranging from 1.1 to 1.6.~\cite{matioli_room-temperature_2015,hirayama_ballistic_1991,tarucha_bend-resistance_1992}.  A decay with two decay constants has previously been observed in semiconductor heterostructures within the ballistic regime~\cite{takagaki_ballistic_1990,chamberlain_electron-boundary_1990,sakamoto_impurity_1991} although the geometry was different and only the region with length scales smaller than the mean free path was considered. In contrast, for all our squares $w/l$~>~1, demonstrating that strong ballistic effects can be observed far outside the region where the device geometry is smaller than the mean free path.
 
Performing the same procedure with the PdCoO$_2$ square (with a longer mean free path of 20~\si{\micro} m) (Fig. S2) and a PtCoO$_2$ square of a significantly lower growth purity and hence with a mean free path of 3~\si{\micro} m results in observed decay rates as a function of $w/l$ within 10\% of the value for the higher-purity PtCoO$_2$ squares (Fig. S3), even though the anisotropy appears at a very different absolute square size in both cases. The difference in A-site cation and curvature of the facets of the Fermi surface which exists between PtCoO$_2$ and PdCoO$_2$ does not seem to significantly affect the decay rate as a function of $w/l$. This implies that, once the overall symmetry of the Fermi surface is determined, this ratio determines the rate of decay. In addition, varying $l$ by changing the temperature also produces double decay behavior consistent with the size dependent studies, providing further evidence that $w/l$ controls this rate (Fig. S4). 

\subsection{Hall voltage}
To complement the bend voltage studies, we also performed a number of Hall voltage measurements, with the configuration $V_H$ = $V_{13,42}$, as shown in the insets to Figs.~4a and b. The Hall voltage at 5~K as a function of magnetic field, normalized by current and sample thickness, is shown in Figs.~4a and b for E1 and D1 respectively, at a number of different side lengths. 

Remarkably, in the diminished orientation, the Hall voltage is finite at zero field and increases as the square size is decreased. For materials with the D$_6$ symmetry of the delafossites within the Ohmic regime, and even within the ballistic regime for materials with a circular Fermi surface, this voltage must be zero at zero field. Here, the combination of the asymmetry of the Fermi surface about the square diagonals and the intrinsic non-locality of the ballistic regime enables non-zero values to be observed for the first time in a material without intrinsic Ohmic conductivity anisotropy. This behaviour can even be controlled by varying the symmetry: in the enhanced orientation, where the Fermi surface orientation is symmetric along the square diagonals, the zero-field voltage remains zero.

 At small magnetic field, unlike in some other ballistic-regime studies~\cite{roukes_quenching_1987}, no quenching (suppression) of the Hall voltage is observed in either orientation; we note that other previous experiments and calculations in this regime have shown a variety of low-field behavior highly dependent on the junction geometry and the level of collimation~\cite{baranger_quenching_1989,baranger_geometrical_1990,chang_quenching_1989,ford_influence_1989}. However, both our measurements show a clear deviation at low field from the linear relationship seen in the Ohmic regime at low temperature and present at high field, most strongly in smaller squares. In addition, the Hall voltage is antisymmetric in magnetic field in the enhanced orientation, but is clearly not antisymmetric in the diminished orientation.

To shed light on the origin of the peaks in the Hall voltage, in Figs.~4c and d we show the same data as in 4a and 4b, but rescaled as a function of $w/r_c$ and offset for clarity. It is clear that the peaks occur at the same value of $w/r_c$, meaning that they stem from commensurability between the side length and cyclotron radius, similar to the peaks seen in the bend voltage. In particular, for both orientations there is evidence of a voltage plateau around $w/r_c=2$, at the same position as a peak in the bend resistance and which grows as the square size decreases. This likely corresponds to the ‘last plateau’ reported in other studies, where the magnetic field ensures the electron trajectories are predominately guided into a contact adjacent to the injection contact~\cite{roukes_quenching_1987,beenakker_billiard_1989,ford_influence_1989}. 

At first sight, the finite Hall voltage at zero field and the pronounced asymmetry in the diminished orientation look like evidence for a violation of the Onsager relations, or even a breaking of time reversal symmetry.  However, neither of these possibilities is true. Whilst the Onsager relations were originally derived for the conductivity in the Ohmic regime, Büttiker famously showed that one equivalent relation, $V_{ab,cd}(B)/I = V_{cd,ab}(-B)/I$, holds true even when transport becomes non-local~\cite{buttiker_symmetry_1988}. We checked this by performing the appropriate measurement combinations in our device.  As seen in Figs. 4e and f, the Büttiker relation is precisely obeyed in our data, even in this highly non-local regime, and even when the symmetry of our device plus Fermi surface combination is so low that the Hall voltage is completely asymmetric in field.  Indeed, our measurements are a ‘text-book’ verification of the Onsager-Büttiker predictions.

\section{Discussion}

Arguably the key feature of our experimental findings is the pronounced effect of setting up a situation in which the overall symmetry of a ballistic device is lowered from the four-fold square symmetry of the device itself, in spite of the fact that we are working with a material whose bulk Ohmic conductivity is isotropic by symmetry.  To better understand the overall pattern of symmetries across our data set, we have performed an analysis using the Landauer-Büttiker multiprobe formula~\cite{buttiker_four-terminal_1986} whilst assuming both current conservation and time-reversal symmetry. This method frames the problem in terms of the transmission probability between pairs of contacts and is a common approach for circular Fermi surface materials~\cite{baranger_classical_1991,hirayama_ballistic_1991,gilbertson_ballistic_2011} but here we lower the symmetry from four-fold to two-fold to reflect the lower symmetry of the hexagonal Fermi surface. Our primary aim in this modelling is to understand the consequences of time reversal symmetry and current conservation for the electronic transport in this lower-symmetry case. Further details of the calculation are given in the supplementary information.

Within this model, the bend resistances, $R_{\mathrm{B}1} = V_{\mathrm{B}1}/I$ and $R_{\mathrm{B}2} = V_{\mathrm{B}2}/I$, are calculated to be:

\begin{align}
R_{\mathrm{B} 1}&=\frac{d}{3 t N} \frac{h}{2 e^{2}} \frac{T_{41} T_{32}-T_{31} T_{42}}{D(T_{41}+T_{21})} \\
R_{\mathrm{B} 2}&=\frac{d}{3 t N} \frac{h}{2 e^{2}} \frac{T_{34} T_{21}-T_{31} T_{42}}{D(T_{41}+T_{21})}
\end{align}

where $T_{ij}$ is the probability of an electron emitted from contact $j$ being absorbed by contact $i$, $N$ is the number of 1D channels at the Fermi energy within each of the contacts, $d$ is the \emph{c}-axis lattice parameter and $D$ is a positive-definite collection of transmission coefficients which is symmetric in the field. With the assumed symmetries and conservation laws, both bend voltages are field-symmetric within this model, as is the case within the data (Fig. S5), even though the electron injection in the diminished orientation is not symmetric with respect to the square diagonal and such symmetry in the transport is therefore not obvious. 
 
In addition, it is clear that the sign of the bend voltage is determined by the relative probability of electron transmission between diagonally linked contacts in Fig.~1 versus those horizontally linked for Bend~1 and vertically linked for Bend~2. In materials with a circular Fermi surface and in the enhanced orientation for our hexagonal Fermi surface, there is a collimated beam of electrons directed along the square diagonal. Therefore, $T_{31}T_{42}$ becomes larger than the first numerator term and the bend voltage is negative. In the limit that horizontal and vertical transmission are equally likely, which is always applicable in these two situations, $T_{41}T_{32} = T_{34}T_{21}$ and the Bend 1 and Bend 2 voltages are equal. 

In the diminished orientation, there is a beam of electrons aligned to the vertical direction, and no such corresponding horizontal or diagonal beam. Therefore, at zero magnetic field, vertical transmission is significantly more probable than horizontal or even diagonal transmission, the $T_{41}T_{32}$ term dominates and R$_{B1}$ is positive.  Conversely, $T_{34}T_{21}$ is smaller than $T_{31}T_{42}$ and $R_{B2}$ is negative. 

For both orientations, in larger magnetic fields, the trajectories are significantly curved by the field, which reduces $T_{31}T_{42}$ and results in a positive voltage. When $w$ is a multiple of the cyclotron diameter, $2r_c$, $T_{41}T_{32}$ and $T_{34}T_{21}$ are large compared to $T_{31}T_{42}$, leading to the transverse focusing peaks in the magnetoresistance.

The Hall resistance calculated in this model, $R_{\mathrm{H}}(B) = V_{\mathrm{H}}/I$ is 
\begin{equation}
R_{\mathrm{H}}(B)=\frac{d}{3 t} \frac{h}{2 e^{2} N} \frac{T_{41}(B)-T_{21}(B)}{D(B)}
\end{equation}

The Hall voltage is therefore, as is logical, determined by the relative probability of transmission to the contact clockwise of the emission contact compared to that anticlockwise. At negative fields, $R_{\mathrm{H}}(B)$ is:
\begin{equation}
R_{\mathrm{H}}(-B)=-\frac{d}{3 t} \frac{h}{2 e^{2} N} \frac{T_{21}(-B)-T_{41}(-B)}{D(B)}
\end{equation}

Even taking into account the time-reversal symmetry that is assumed in the model, $R_{\mathrm{H}}(-B)$ is only equal to $-R_{\mathrm{H}}(B)$, meaning the Hall voltage is anti-symmetric, if $T_{41}(B) = T_{21}(-B)$, which is equivalent to stating that the transmission coefficients for clockwise and anticlockwise transmission are related by inverting the field. This antisymmetry condition is only satisfied if there is a line of symmetry of the Fermi surface along the diagonal of the square. For the enhanced orientation, this is the case, but it is not true for the diminished orientation, explaining the lack of antisymmetry in the Hall data in Fig.~4b and, in addition, demonstrating that asymmetry is possible without a violation of time-reversal symmetry. Furthermore, in all cases, as is shown in the supplementary information, the Büttiker-Onsager relations are adhered to within the model.
 
The analysis therefore shows that the novel anisotropy observed in the ballistic regime junctions can be explained by an anisotropy in the probability of transmission of electrons between contacts, stemming from the hexagonal Fermi surface. This can result in anisotropy in the bend voltage and a lack of antisymmetry of the Hall voltage in an applied magnetic field. Important constraints, from both current conservation and time-reversal symmetry, lead to the bend voltage being symmetric in the field and the Hall resistance obeying the Onsager relations, in their non-local form. Crucially, pronounced asymmetry in the Hall signal in the diminished orientation does not violate any of these fundamental symmetries, demonstrating their continued existence even in this highly non-local regime. 

Although the main aim of our Landauer-Büttiker analysis was to understand the overall patterns and symmetries seen in our data, it can also provide quantitative insight into the scale of the effects that we observe.  If a system is two-dimensional and fully ballistic, the minimum two-terminal resistance $R_{\mathrm{2 T}}$ for any given configuration is

\begin{equation}
R_{\mathrm{2 T}}=\frac{h}{2 e^{2} N} \frac{1}{N_{\mathrm{L}}}
\end{equation}

in which $N$ is the number of conductance channels in the current contacts and $N_{\mathrm{L}} = 3t/d$ is the number of Pt/Pd layers in parallel. For contacts modelled with square well potentials, $N = k_Fc/\pi$ where $c$ is the width of the contact channel~\cite{van_houten_chapter_1992}. In typical experiments on semiconductor squares and crosses, there is a single conducting layer and $N$ is usually of order 10 or less. In contrast, for the devices and contact sizes used in the current work, the large $k_F$ in our delafossite devices ensures that $N\sim 12000$ per Pt/Pd layer and there are $N_L\sim 3000$ layers in parallel, possibly accounting for the lack of conductance fluctuations noted above in relation to the data shown in Fig. 3. This leads to $R_{\mathrm{2T}} \cong 0.3$m$\Omega$.

In our experiments on these squares we have not directly measured this two-point resistance, but our measured bend and Hall voltages and the model grant the opportunity to benchmark our results against values predicted in the ballistic limit within the Landauer-Büttiker quantum transport formalism. Within this formalism, as shown in the supplementary information, if all of the electron trajectories are presumed to be guided into an adjacent contact by the magnetic field, as occurs on the last plateau around $w/r_c = 2$, $R_{\mathrm{H}} = R_{\mathrm{2T}}$. As can be seen in Fig. 4a, for the 15~\si{\micro} m enhanced orientation square, $R_{\mathrm{H}}t \cong 0.4 $~m~\si{\Omega\micro}m on this plateau, leading to $R_{\mathrm{H}}= 0.16$~m$\Omega$, within a factor of two of the fully ballistic prediction for $R_{\mathrm{2T}}$. Fully quantitative analysis of the bend resistance data would require assumptions about transmission coefficients.  However, we note that the difference between the bend voltages in the diminished orientation is approximately double that between the 7~\si{\micro} m and 35~\si{\micro} m enhanced orientation squares, providing an estimate for a ballistic bend resistance $R_{\mathrm{B}} \cong 0.12$~m$\Omega$, within a factor of 2.5 of $R_{\mathrm{2T}}$. 

Our squares are not fully in the ballistic limit, but the above estimates further confirm that the ballistic effects predicted by the Landauer-Büttiker quantum transport formalism play a major role in the behavior we observe in the size-restricted squares, even though all of our squares are larger than the typical 5~\si{\micro} m low-temperature mean free path.  This can also be seen by examination of Fig.~3d.  The form of signal decay is exp$(-bL/l)$ where $L$ is a sample length scale and $l$ the mean free path.  Definitions and prefactors will of course vary with precise device geometry, but the exponential can sensibly be extrapolated to the true ballistic limit in which the mean free path is much longer than the device size and $w/l$ approaches zero.  For the diminished orientation experiments summarized in Fig.~3d this shows that our smallest squares allow us to observe approximately 50\% of the fully ballistic signal, fully consistent with the above ballistic resistance estimates.

Our findings, therefore, provide an empirical verification that transport within a metal microstructure with >~$10^4$ quantum channels per layer can still be well described using Landauer-Büttiker theory, despite the significant difference in energy scales compared to semiconductor devices.  We can further profit from the strong anisotropies of our system to follow the signatures of ballistic physics into a regime which one might intuitively have expected to be out of reach.  As seen in Fig.~3c, there remains a difference between the two bend voltages even at a side length of 95~\si{\micro} m, more than 15 times the mean free path in this PtCoO$_2$ square. This is partly due to the sensitivity of having a difference measurement with which to pick out the ballistic signature, but we speculate that there may be other physics at play in this regime (that of the slower decay in Fig.~3d) as well.  In materials with a circular Fermi surface, mean free paths extracted from ballistic phenomena are often shorter than those determined from the resistivity. This is often attributed to small-angle scattering. This scattering contributes little to the resistivity, but significantly affects the electron trajectories and therefore ballistic regime behavior. In PtCoO$_2$, however, small angle scattering frequently does not appreciably change the zero-field trajectory as the post-scattering state will typically be on the same facet of the hexagonal Fermi surface. This may mean that bulk scattering is less effective in suppressing ballistic effects than in materials with a circular Fermi surface, leading to effects persisting over longer length scales and smaller decay constants; our observations strongly motivate future theoretical work on this issue.

In conclusion, we have shown here how some effects previously assumed to lie exclusively in the realm of ultra-pure semiconductor devices can be observed in sufficiently pure two-dimensional metals.  Arguably more interestingly still, we have shown how the existence of a strongly faceted Fermi surface leads to phenomena not previously observed in any system, opening the possibility of new regimes of mesoscopic physics.

\subsection*{Acknowledgments}
	We are pleased to acknowledge useful discussions with L.W. Molenkamp and B. Schmidt. 

\subsection*{Funding}
Max Planck Society. 
Engineering and Physical Science Research Council PhD studentship support via grant EP/L015110/1 (PHM,MDB).
Deutsche Forschungsgemeinschaft Cluster of Excellence ct.qmat EXC 2147, project-id 390858490. 
European Research Council (ERC) under the European Union’s Horizon 2020 research and innovation programme (MiTopMat - grant agreement No 715730) (PJWM,CP).

 \subsection*{Materials and methods}
 \subsection*{Crystal growth}
Single crystals of PdCoO$_2$ (PtCoO$_2$) were grown in an evacuated quartz ampule with a mixture of PdCl$_2$ (PtCl$_2$) and CoO by the following methathetical reaction:
PdCl$_2$ (PtCl$_2$) + 2CoO $\rightarrow$ 2PdCoO$_2$ (PtCoO$_2$) + CoCl$_2$. The ampule was heated at 1000~$^{\circ}$C for 12 hours and stayed at 700-750~$^{\circ}$C for 5 (20) days. In order to remove CoCl$_2$, the resultant product was washed with distilled water and ethanol.

\subsection*{Device fabrication}
The FIB microstructuring was performed using a Ga-ion FEI Helios NanoLab FIB at a 30~keV acceleration voltage. Each crystal was placed in an epoxy droplet with a 200~nm gold layer then sputtered over the surface to provide electrical contacts. Ion beam currents of either 9.1~nA or 21~nA were used to sculpt the initial square devices, with lower currents of 0.79~nA or 2.5~nA used for sidewall polishing and later size-reduction steps. To prevent issues with current inhomogeneity due to the large resistivity anisotropy of the delafossite crystals and the top injection of current, long meanders at least 150~\si{\micro}m in length were structured from the crystal leading to each square contact. These meanders ensured the current penetrated the full thickness of the crystal before reaching the square. The gold layer was removed from the surface of the square and the meanders and separated into four electrical contact pads by FIB irradiation. 

\subsection*{Transport measurements}
All electrical transport measurements were made using a bespoke low-noise probe placed within a Quantum Design Physical Property Measurement System. Standard ac lock-in techniques were used for the voltage measurements at a frequency of 123~Hz and with currents between 3 and 9~mA using a bespoke dual end current source providing high common mode rejection and a Synktek MCL1-540 multichannel lock-in. The transverse magnetic field in the bend magnetoresistance and Hall measurements ranged between -9~T and +9~T.

\bibliographystyle{imacrev}

\bibliography{squares_arxiv_postsubmit.bib} 
\clearpage
\begin{figure}[t!]
	\centering
	\includegraphics[width=0.8\textwidth]{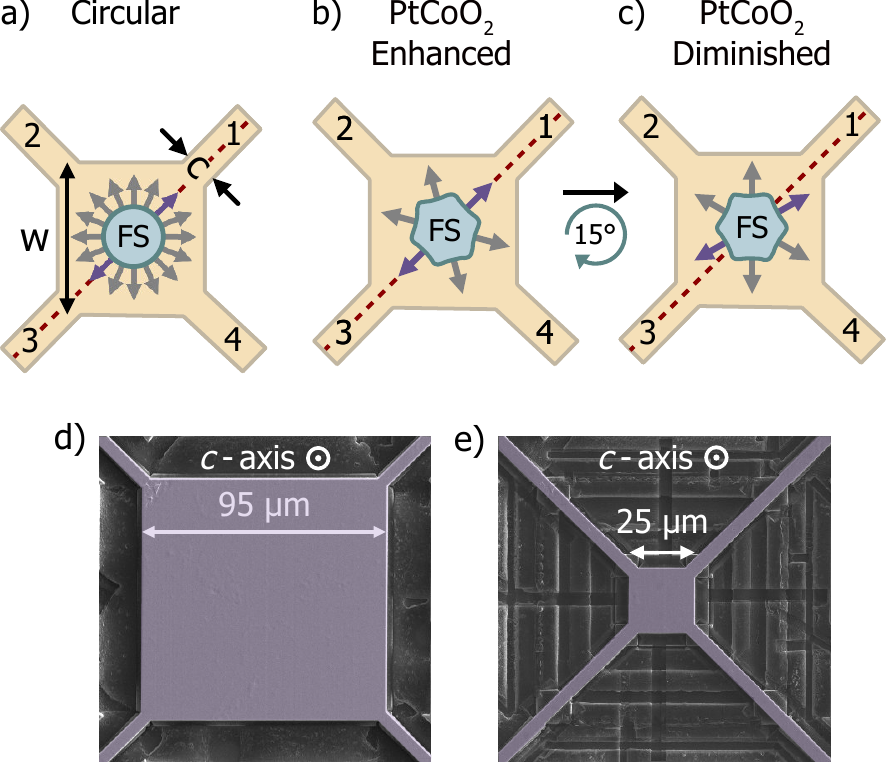}
	\caption{Design and orientation of square microstructures. (a-c) Diagrams of the orientation of the Fermi surface (FS) relative to a square of side length $w$ and contact width $c$ for (a) a circular FS and the PtCoO$_2$ Fermi surface with (b) an enhanced orientation where the Fermi surface is symmetric about the square diagonals, designed to increase diagonal transport and (c) a diminished orientation where the Fermi surface is not symmetric about the diagonals and where the diagonal transport is reduced. (d-e) False color scanning electron microscope images, at the same scale, of a PtCoO$_2$ square structured using a focused ion beam (FIB) with (d) side length 95~\si{\micro}m  and (e) with the side length reduced to 25~\si{\micro}m using the FIB.}
	\label{fig:fig1}
\end{figure}

\clearpage

\begin{figure}[t]
	\centering
	\includegraphics[width=0.8\textwidth]{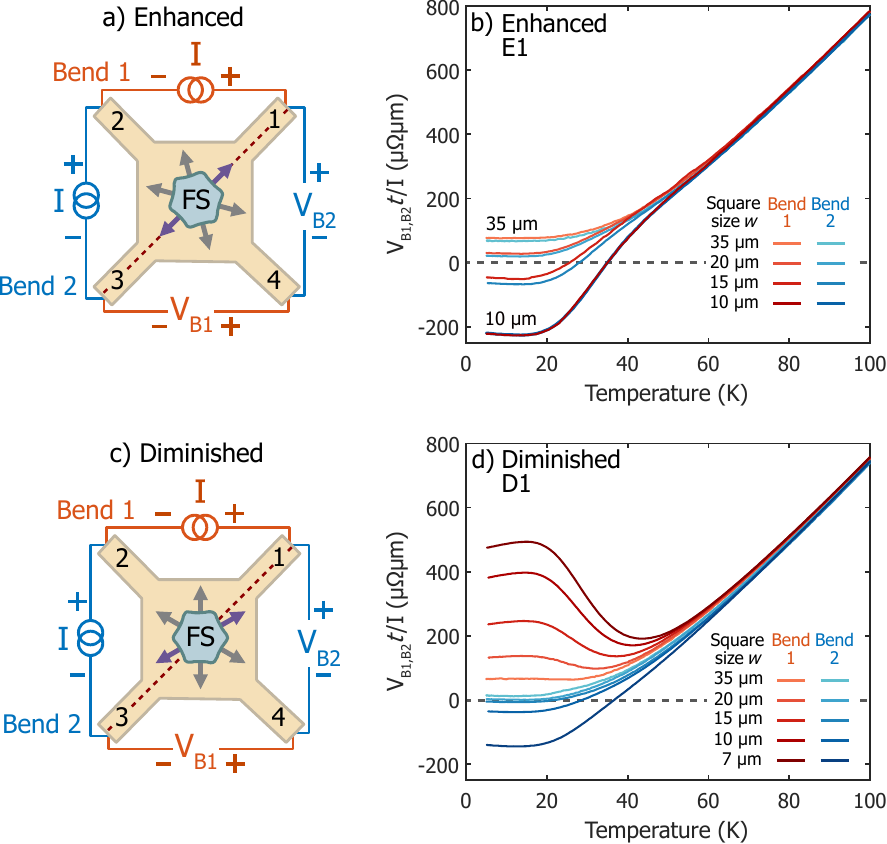}
	\caption{Temperature dependence of bend resistance. (a) Diagram of the contact configurations used to measure the bend voltages, Bend 1 ($V_{\mathrm{B}1}$) and Bend 2 ($V_{\mathrm{B}2}$), shown on an enhanced orientation square. (b) The temperature dependence below 100~K of $V_{\mathrm{B}1,2}t/I$, where $t$ is the square thickness and $I$ is the current, for an enhanced orientation square, E1, with side lengths between 10~\si{\micro}m and 35~\si{\micro}m. Although this quantity has the units of resistivity it only gives the true resistivity in the Ohmic regime, and becomes strongly size and orientation-dependent in the non-local regime. Its primary function is to normalize variations in thickness or measurement current between squares. (c) The contact configurations for both bend voltage measurements shown on a diminished orientation square. (d) The temperature dependence below 100~K of $V_{\mathrm{B}1,2}t/I$ for a diminished orientation square, D1.}
	\label{fig:fig2}
\end{figure}

\clearpage

\begin{figure}[t]
	\centering
	\includegraphics[width=\textwidth]{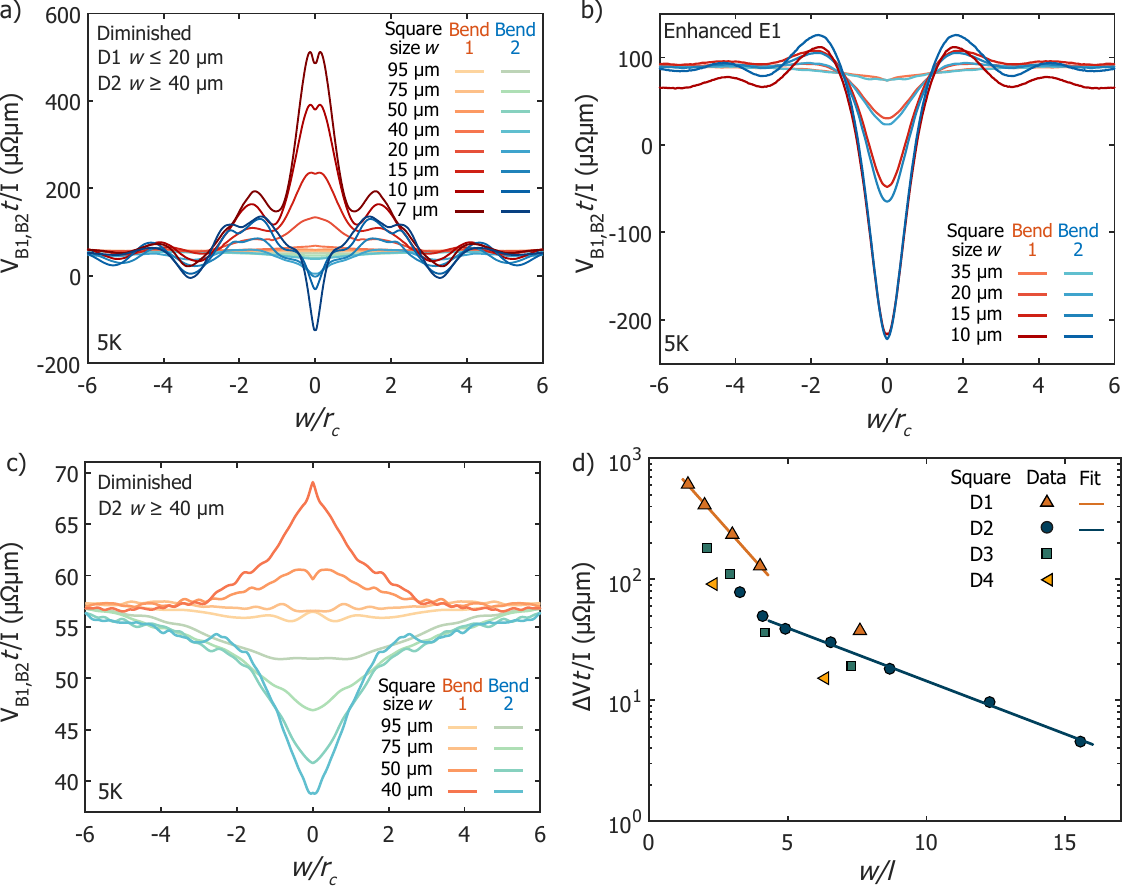}
	\caption{Magnetic field dependence of bend resistance.  The variation at 5~K of $V_{\mathrm{B}1,2}t/I$ with $w/r_c$, a quantity proportional to magnetic field, within (a) two diminished orientation squares, D1 and D2, at side lengths from 7~\si{\micro} m to 20~\si{\micro} m and 40~\si{\micro} m to 95~\si{\micro} m respectively (b) the enhanced orientation square E1 for side lengths between 10~\si{\micro} m and 35~\si{\micro} m and (c) only the larger diminished orientation D2 for side lengths 40~\si{\micro} m and larger. The magnetic field was applied parallel to the \emph{c}-axis and therefore perpendicular to the square. Negative values of $w/r_c$ indicate negative magnetic fields. (d) The decay of $\Delta Vt/I$, where $\Delta V=V_{\mathrm{B}1}-V_{\mathrm{B}2}$ at zero field, as a function of the ratio of the square side length $w$ to the mean free path $l$ for four diminished orientation squares, D1 to D4.}
	\label{fig:fig3}
\end{figure}

\clearpage

\begin{figure}[t]
	\centering
	\includegraphics[width=\textwidth]{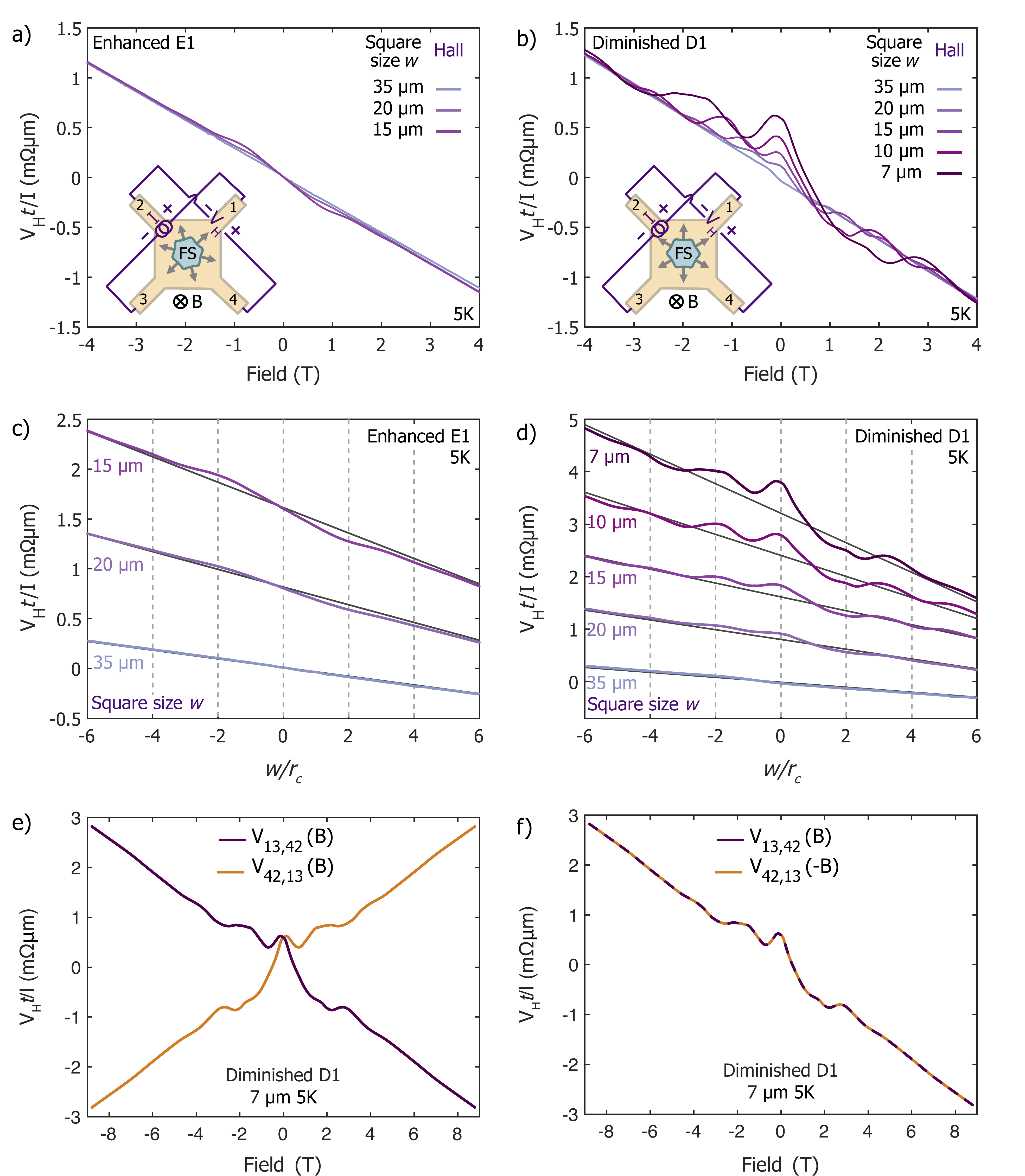}
	\caption{Hall voltage. (a and b) The variation at 5~K of $V_{\mathrm{H}}t/I$, where $V_{\mathrm{H}}$ is the Hall voltage, with magnetic field, measured with the configuration shown in the inset diagrams for (a) the enhanced orientation square E1 at side lengths between 15~\si{\micro} m and 35~\si{\micro} m and (b) the diminished orientation square D1 at square side lengths between 7~\si{\micro} m and 35~\si{\micro} m. (c and d) The data from a) and b) respectively rescaled as a function of $w/r_c$ with an offset of 0.8~m$\Omega$\si{\micro} m for clarity and a linear fit to data at high field for each side length. (e) The data for the 7~\si{\micro} m side length of square D1 ($V_{\mathrm{H}}=V_{13,42}$) alongside a measurement for the same square but with the current and voltage contacts switched ($V_{42,13}$).(f) The data from e) but with the data for the second contact configuration flipped in the field ($B$) to demonstrate a perfect agreement with the Onsager relations.}
	\label{fig:fig4}
\end{figure}

\clearpage

\renewcommand\thefigure{S\arabic{figure}}    
\setcounter{figure}{0}

\section*{Supplementary information}
\subsection*{Landauer-Büttiker model}

The Büttiker multichannel equation states that, in a multicontact measurement at zero temperature, $I_i$, the net current at contact $i$, is
\begin{equation}
I_{\mathrm{i}}=\frac{2 e N}{h}\left[\left(1-R_{\mathrm{i}}\left(E_{\mathrm{F}}\right)\right) \mu_{\mathrm{i}}-\sum_{j, j \neq i} T_{\mathrm{ij}}\left(E_{\mathrm{F}}\right) \mu_{\mathrm{j}} \right]
\end{equation}\\
where $\mu_i$ is the chemical potential at contact $i$, $N = k_Fc$, where $k_F$ is the Fermi wavelength, is the number of 1D-modes at the Fermi energy in each contact of width $c$, $ T_{ij}(E_{\mathrm{F}})$ is the probability an electron with the Fermi energy $(E_{\mathrm{F}})$ emitted by contact $j$ is absorbed by contact $i$, and $R_i$ is the probability of reflection of the electron back into the emission contact~\cite{buttiker_four-terminal_1986}. For brevity, in the discussion below and in the main text, $T_{ij}=T_{ij}(E_{\mathrm{F}})$. 

In Figure 1, for both the diminished and enhanced orientation, contact 1 is equivalent to contact 3 and contact 2 is equivalent to contact 4 in terms of transmission probabilities to adjacent and diagonal contacts, for example $T_{32} = T_{14}$ and $T_{34} = T_{12}$. Considering this, the above equation can be written as 
\begin{equation}
\left[\begin{array}{c}
I_{1} \\
I_{2} \\
I_{3} \\
I_{4}
\end{array}\right]=\frac{2 e N}{h}\left[\begin{array}{cccc}
T_{A} & -T_{34} & -T_{31} & -T_{32} \\
-T_{21} & T_{B} & -T_{41} & -T_{42} \\
-T_{31} & -T_{32} & T_{A} & -T_{34} \\
-T_{41} & -T_{42} & -T_{21} & T_{B}
\end{array}\right]\left[\begin{array}{c}
\mu_{1} \\
\mu_{2} \\
\mu_{3} \\
\mu_{4}
\end{array}\right] \label{eq:matrix}
\end{equation}\\
where $T_{\mathrm{A}} = 1-R_1 = \sum_{j,j\neq  1}T_{j1} = 1-R_3$ and $T_{\mathrm{B}} = 1-R_2 = \sum_{j,j\neq 2}T_{j2}=1-R_4$ as required by probability conservation. Current conservation imposes that $\sum_jT_{ij}=\sum_i T_{ij}$, hence $T_{21}+T_{41}=T_{34}+T_{32}$, and time reversal symmetry requires $T_{ij}(B)=T_{ji}(-B)$~\cite{datta_electronic_1997}. 

In the enhanced orientation, the Fermi surface, and therefore the possible directions of electron transmission from a contact, are symmetric about the square diagonals. This ensures that the transmission probabilities to the clockwise and anticlockwise adjacent contacts are symmetric in the magnetic field, for example, T$_{21}$(B)=T$_{41}$(-B). The diminished orientation instead has symmetry lines of the Fermi surface and the electron transmission parallel to the square edges. However, the consequent  symmetries in the transmission coefficients, such as T$_{21}$(B) = T$_{43}$(B) = T$_{12}$(-B) = T$_{34}$(-B), are no different than those already imposed by the time reversal symmetry and the two pairs of square contacts with equal transmission coefficients discussed above, which exist in both orientations. This therefore does not impose an additional symmetry on the transmission coefficient matrix.

To solve Equation~\ref{eq:matrix}, the net current at the voltage contacts is set to zero and the value at the current terminal set to $\pm I$ at the positive/negative current contact. The voltage is then determined from the chemical potential difference from the following equation, 

\begin{equation}
\frac{V_{i j, k l}}{I}=\frac{\mu_{i}-\mu_{j}}{e I} \frac{d}{3 t}
\end{equation}\\
where the $d/3t = 1/N_L$ factor is the inverse of the number of parallel conducting Pt/Pd layers in our delafossite devices, $N_L$.

As discussed in the main text, the bend resistances deduced from this model are

\begin{equation}
\frac{V_{B 1}}{I}=\frac{V_{12,43}}{I}=\frac{d}{3 t} \frac{h}{2 e^{2} N} \frac{T_{41} T_{32}-T_{31} T_{42}}{D\left(T_{41}+T_{21}\right)}
\end{equation}\\
and
\begin{equation}
\frac{V_{B 2}}{I}=\frac{V_{23,14}}{I}=\frac{d}{3 t} \frac{h}{2 e^{2} N} \frac{T_{34} T_{21}-T_{31} T_{42}}{D\left(T_{41}+T_{21}\right)}
\end{equation}\\
where $D=T_{41} T_{34}+T_{21} T_{32}+T_{34} T_{31}+T_{32} T_{31}+T_{41} T_{42}+T_{21} T_{42}+2 T_{31} T_{42}$, a quantity symmetric in the field. Imposing the symmetries and conservation laws discussed above constrains both bend resistances to be symmetric in magnetic field.

The Hall resistance 

\begin{equation}
\frac{V_{H}}{I}=\frac{V_{13,42}}{I}=\frac{d}{3 t} \frac{h}{2 e^{2} N} \frac{T_{41}-T_{21}}{D}
\end{equation}\\
which, as discussed in the main text, is only antisymmetric in field if the Fermi surface is symmetric about the square diagonal. 

However, considering the exchange of current and voltage contacts,

\begin{equation}
\frac{V_{H 2}}{I}=\frac{V_{42,13}}{I}=\frac{d}{3 t} \frac{h}{2 e^{2} N} \frac{T_{32}-T_{34}}{D}
\end{equation}\\
which, using time-reversal symmetry, obeys the Büttiker-Onsager reciprocal relations $V_{\mathrm{H1}}(B)/I=V_{\mathrm{H1}}(-B)/I$ regardless of the presence of four-fold symmetry. 

When guiding dominates and all electrons trajectories lead to an adjacent contact and assuming a field direction causing clockwise motion, $T_{41}=T_{34}=1$ with all other transmission coefficients equal to zero. In this limit, $D=1$ and therefore
\begin{equation}
\frac{V_{H}}{I}=\frac{d}{3 t} \frac{h}{2 e^{2} N}=R_{2 \mathrm{T}}
\end{equation}\\
where $R_{2\mathrm{T}}$ is the ballistic two-terminal resistance discussed in the main text. 

\clearpage
\subsection*{Current dependence}
\begin{figure}[h]
	\centering
	\includegraphics[width=0.6\textwidth]{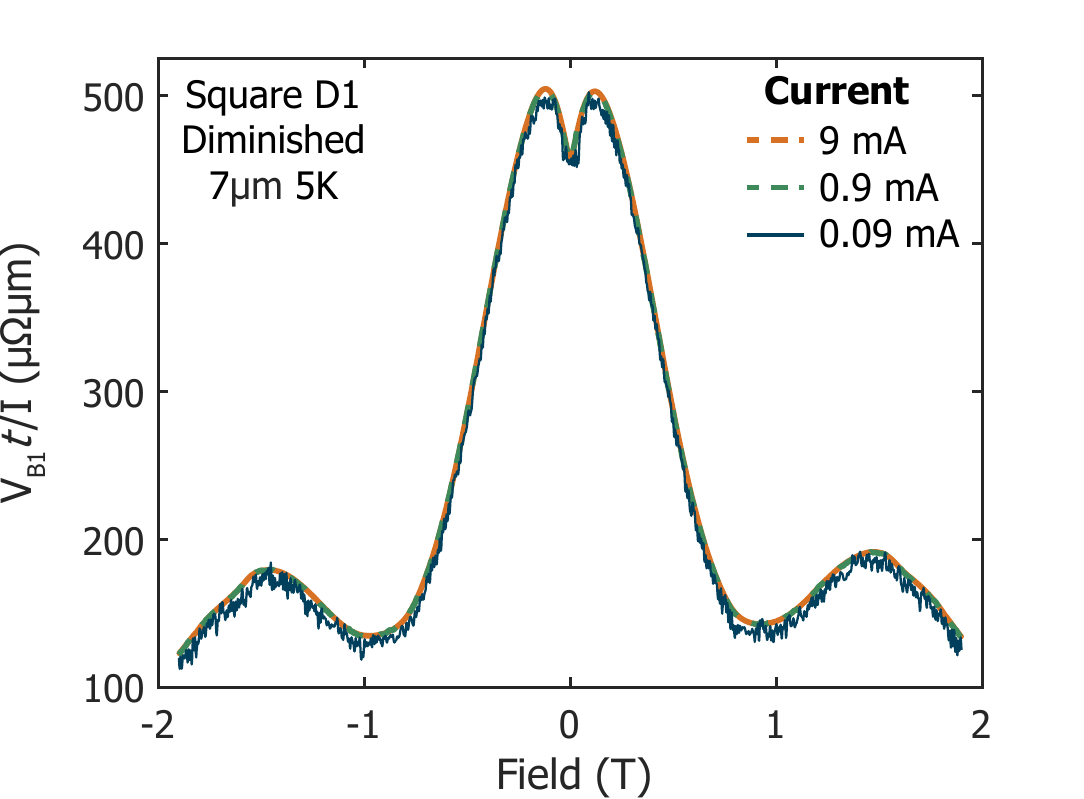}
	\caption{Current dependence. The change in $V_{\mathrm{B1}}t/I$ as a function of the magnetic field in D1 with a 7~\si{\micro} m side length at three different values of current. }
	\label{fig:figS1}
\end{figure}

\subsection*{Behaviour in a PdCoO$_2$ square}
\begin{figure}[h]
	\centering
	\includegraphics[width=\textwidth]{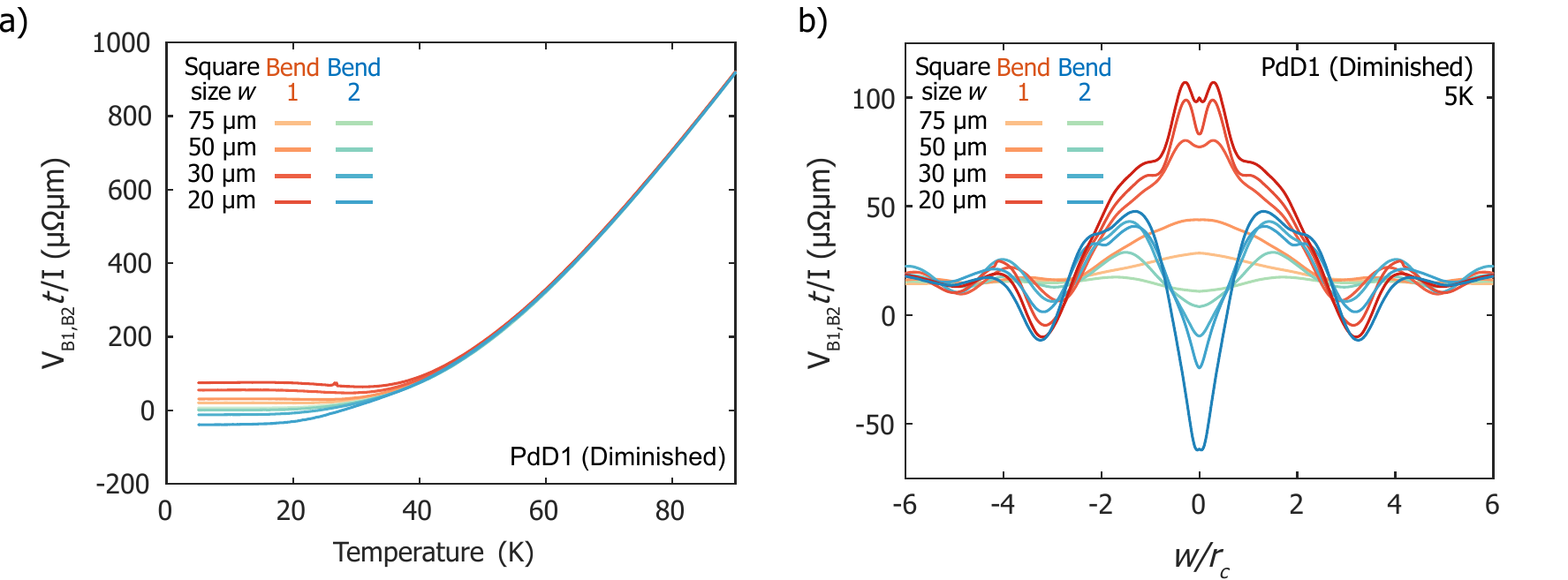}
	\caption{a) The temperature dependence below 100K of $V_{\mathrm{B1,B2}}t/I$, for a PdCoO$_2$ diminished orientation square, PdD1, measured at side lengths between 20~\si{\micro} m and 75~\si{\micro} m. b) The variation at 5K of $V_{\mathrm{B1,B2}}t/I$ with $w/r_c$ for the same square and side lengths.}
	\label{fig:figS2}
\end{figure}

\clearpage

\subsection*{Decay of anisotropy in PdCoO$_2$ square and lower-purity PtCoO$_2$ square}
\begin{figure}[h]
	\centering
	\includegraphics[width=0.6\textwidth]{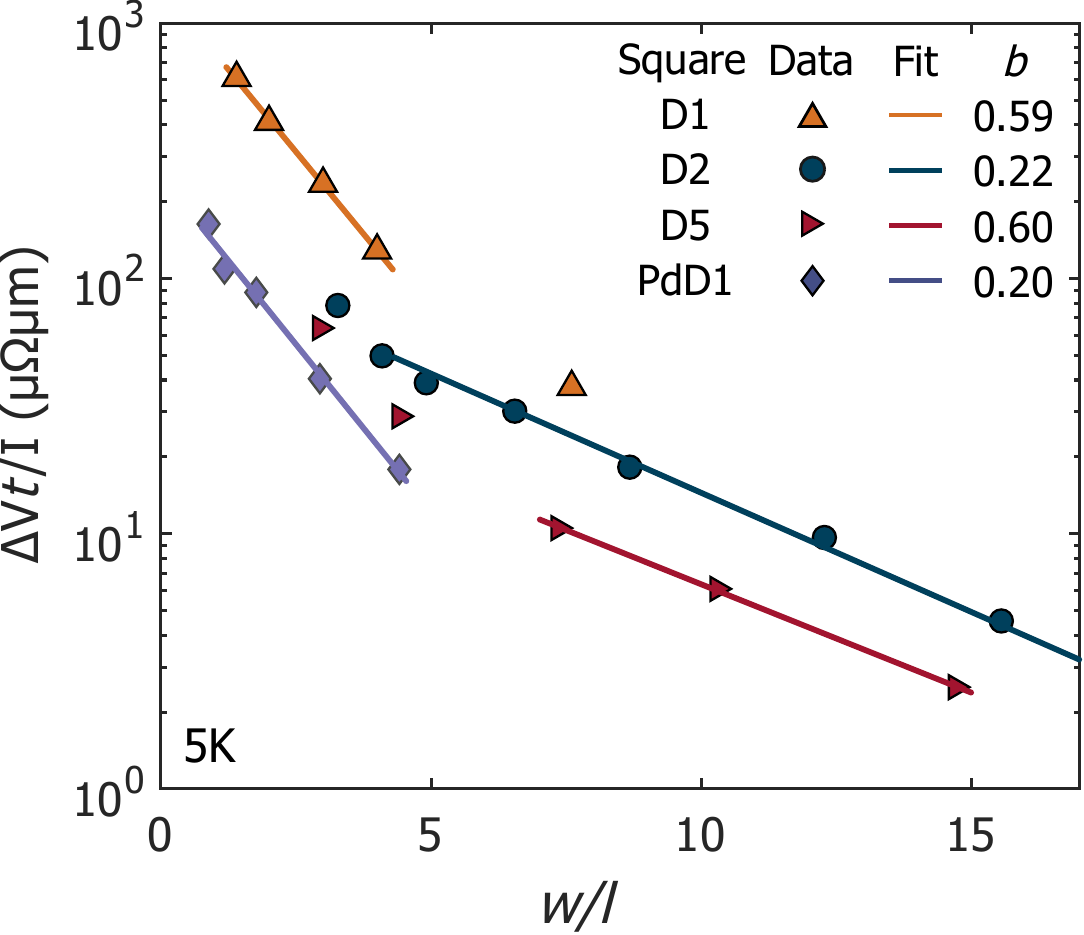}
	\caption{The decay of $\Delta Vt/I$, where $\Delta V=V_{\mathrm{B1}}-V_{\mathrm{B2}}$, as a function of $w/l$ for four diminished orientation PtCoO$_2$ squares, including the lower-purity D5, and the diminished orientation PdCoO$_2$ square, PdD1, alongside exponential fits $\Delta Vt/I=A \mathrm{exp}(-bw/l)$ in either the rapid and slow decay regions identified in Fig. 4 in the main text. }
	\label{fig:figS3}
\end{figure}

\subsection*{Temperature dependence}
\begin{figure}[h]
	\centering
	\includegraphics[width=\textwidth]{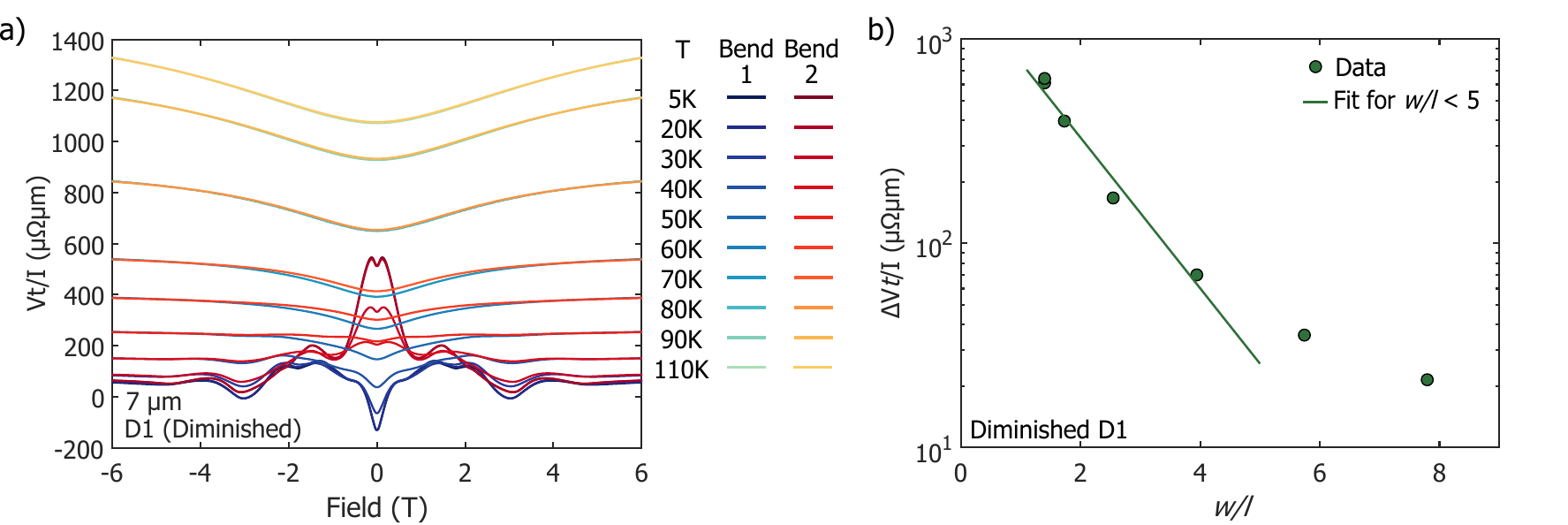}
	\caption{The magnetic field dependence of $V_{\mathrm{B1,B2}}t/I$ at temperatures between 5K and 110K for D1 when $w = 7$\si{\micro} m. (B) The decay of $\Delta Vt/I$, where $\Delta V=V_{\mathrm{B1}}-V_{\mathrm{B2}}$, as a function of $w/l$ for D1, where this ratio is varied by changing the temperature whilst maintaining $w = 7$\si{\micro} m, alongside an exponential fit $\Delta Vt/I=A \mathrm{exp}(-bw/l)$  with $b=0.8\pm0.2$ for the data points with $w/l < 5$.  }
	\label{fig:figS4}
\end{figure}
\clearpage

\subsection*{Symmetry of signal in magnetic field}
\begin{figure}[h]
	\centering
	\includegraphics[width=0.7\textwidth]{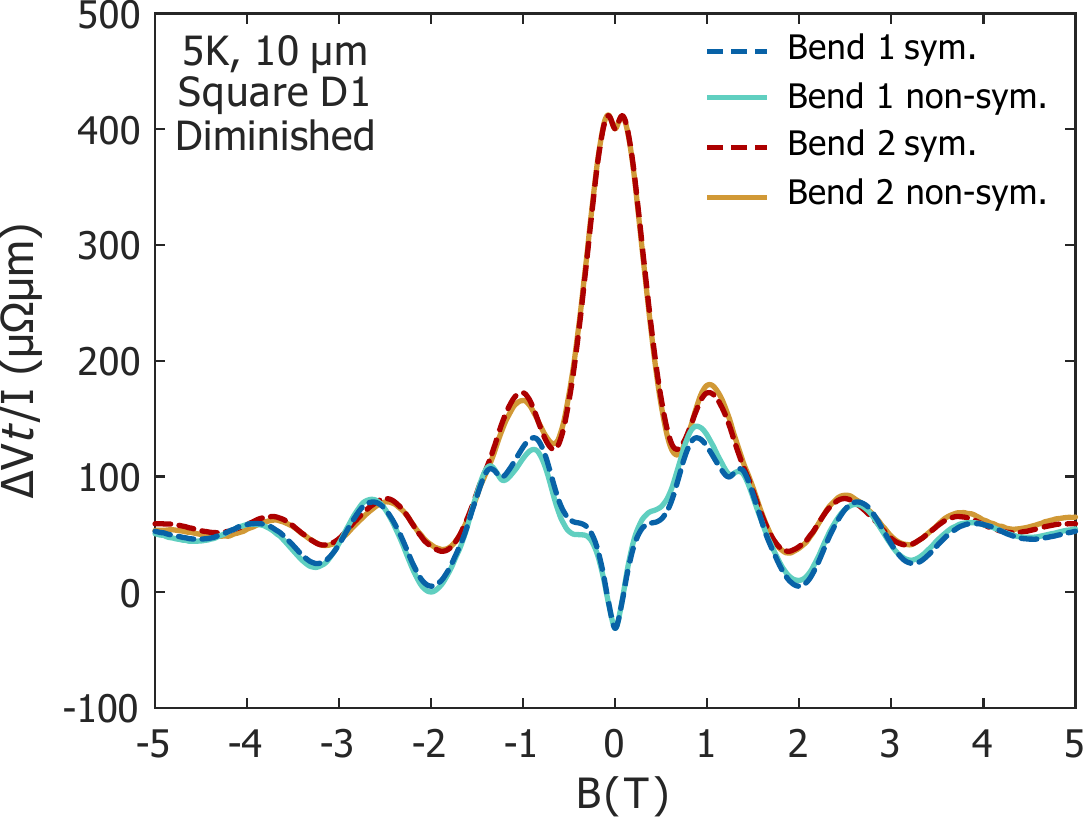}
	\caption{The magnetotransport in D1 at 5K and with a 10~\si{\micro} m side length for both bend voltage measurements and with (sym.) and without (non-sym.) symmetrization of the data in the magnetic field.}
	\label{fig:figS5}
\end{figure}
\end{document}